\documentstyle[aps,preprint]{revtex}
\begin{document}
\draft
\preprint{To be published in PRL}
\title{Strong-Pinning Effects in Low-Temperature Creep:
Charge-Density Waves in TaS$_3$ }
\author{S.~V.~Zaitsev-Zotov\\}
\address{Institute of Radioengineering and Electronics of
Russian Academy of Sciences, Mokhovaya 11,103907 Moscow,
Russia.\\}
\author{G.~Remenyi and P.~Monceau\\}
\address{Centre de Recherches sur les Tr\`es Bases Temp\'eratures, 
CNRS, 25, Avenue des Martyrs, BP 166 X,
38042 Grenoble C\'edex, France\\}
\date{\today}
\maketitle
\begin{abstract}
Nonlinear conduction in the quasi-one dimensional conductor o-TaS$_3$ 
has been studied in the low-temperature region down to 30 mK. It 
was found that at temperatures below a few Kelvins the 
current-voltage (I-V) characteristics consist of several branches. 
The temperature evolution of the I-V curve proceeds through sequential 
freezing-out of the branches. The origin of each
branch is attributed to a particular strong pinning impurity type.
 Similar behavior is expected for other physical 
systems with collective transport (spin-density waves, Wigner
crystals, vortex lattices in type-II superconductors {\it etc.})
in the presence of strong pinning centers. 

\end{abstract}

Collective transport may be observed in
many physical systems such as vortex lattices in 
type II superconductors, electronic crystals like Wigner crystals
and charge- and spin-density waves (CDW and SDW) in low-dimensional 
conductors, and others, as a response to an 
external force coupled with the respective order parameter 
\cite{review,Natterman}. The motion rate is 
slow down by interaction with pinning centers (impurities and
imperfections) and depends also on energy dissipation, fluctuations, 
and other factors.
In general, two limiting types of motion can be considered. 
In the limit of a large driving force the motion rate is
controlled mostly by dissipation and fluctuations can 
be neglected. This regime corresponds to sliding of the 
spin- or charge-density waves and Wigner crystals, 
or to the flux-flow regime for vortex motion 
in type-II superconductors. In the limit of a small
driving force the pinning barriers for motion are large. 
Motion occurs owing to rare activated overcoming 
of, or quantum tunneling through, pinning barriers. As a result, 
dissipation does not dominate any more, and
fluctuations become the main factor determining the motion rate.
This fluctuation-dominating regime of motion is known 
as the creep regime.

A quasi-one-dimensional conductor with a CDW, 
like TaS$_3$, K$_{0.3}$MoO$_3$ {\it etc.} \cite{review}, can be 
considered as a model system which reveals the general properties 
of collective transport.
At relatively high temperatures application of an
electric field, $E$, above the threshold one, $E_T$, leads to the
nonlinear conduction associated with sliding of the CDW. The 
general properties of the sliding regime are known in detail and 
are well understood in the framework of models considering the 
CDW as an elastic medium pinned by impurities \cite{review}. At  
$E<E_T$  the CDW is pinned \cite{nocreep} ,
and the conduction
is due to normal carriers (electrons and holes) excited over 
the Peierls gap 
$2\Delta$ and providing the activated temperature dependence of 
the linear conductivity
$\sigma_{\parallel ,\perp}\propto \exp(-\Delta/T)$, for both 
longitudinal,
$\sigma_\parallel$, and transverse, $\sigma _\perp$, components 
of the conductivity tensor.

The low-temperature linear conduction indicates a contribution of a 
new conduction mechanism which increases $\sigma_\parallel$, but 
has no effect on $\sigma_\perp$. In the particular case of 
orthorhombic TaS$_3$ this contribution starts around $T=60-80$ K
\cite{Tako}, that is about one third of the Peierls transition 
temperature. The picture of the low-temperature nonlinear
conduction is more complex than at high temperatures and includes 
an additional weak nonlinearity which develops below $E_T$ 
\cite{Tako}. $E_T$ grows by two-three orders of
magnitude with lowering temperature down to $T = 4.2$ K, where it 
reaches the value of $10^2-10^3$ V/cm \cite{INM}. 
Lowering temperature makes nonlinearity at
$E<E_T$ more and more pronounced and leads finally to strongly 
nonlinear I-V curves $I\propto (E-E_T)^\alpha$, with the exponent 
$\alpha =15$ at $T=4.2$ K \cite{INM}. Alternatively, the I-V curves 
can also be fitted by the exponential law 
$I\propto \exp \left[ -(V_0/V)^\beta \right]$, 
with $\beta \sim 1-2$ \cite{zzprl}.
The activation energies and other details of the low-temperature 
behavior have a large scatter and vary from sample to sample 
taken even from the same batch \cite{INM}. Many features of the 
low-temperature transport properties of quasi-one dimensional 
conductors like the absence of scaling of nonlinear conduction with 
the linear one, electric-field dependent activation energy, the 
exponential shape of I-V 
curves, temperature-dependent dielectric constant {\it etc.} have 
their natural explanation in terms of CDW creep 
\cite{zzprl,Larkin,LB,Baklanov,Ovchinnikov}. 

In this Letter we present data on the low-temperature 
nonlinear conduction of TaS$_3$ studied in the temperature range 
extended down to 30 mK. It has been found that at sufficiently low 
temperatures the I-V curves consist of several branches. The 
results are described in terms of creep of the CDW pinned by 
strong-pinning centers.


We studied the low-temperature nonlinear conduction of four 
TaS$_3$ single crystals with different impurity contents in the 
temperature region down to 30 mK. Two crystals (referred
hereafter as pure crystals) have the threshold field $E_T$ for the 
onset of the nonlinear conduction below 1 V/cm at 100 K, and four 
others (impure crystals) have $E_T$ above 10 V/cm. In addition, two 
other impure crystals where measured at $T > 1.5$ K.

The current-voltage characteristics were measured by the 
two-contact technique which is the only reliable technique for the 
low-temperature region, where the linear sample resistance 
exceeds $10^{15}$ Ohms. To reduce a possible contribution of 
contacts we studied relatively long samples with typical lengths 
around 5 mm. 
The cross-sectional area of the studied crystals was $10^2-10^3$~$\mu$m$^2$.
Ohmic contacts were prepared by either vacuum deposition
of indium, or cold soldering by indium. 
The shape of the I-V curves was found to be independent of the contact 
preparation technique. 

The measurements were done using a DC voltage-control 
technique, and a Keithley 617 electrometer for current 
measurements. 
 The data were collected from the largest to smallest 
voltage values.
 The delay between applying/changing the voltage 
and taking a current reading was around 1 s at relatively high 
currents $I>10^{-10}$ A, and reached 30 s for currents below 
$10^{-13} $ A. From our experience such a duration was enough for 
the polarization current to fall below the noise level of our setup 
($10^{-14} - 10^{-15}$ A). With such a delay the shape of the measured 
I-V curves was reproducible and independent of the direction of 
the voltage sweep.

The experiments below 1 K were done in a $^3$He-$^4$He 
dilution refrigerator designed especially for the low-noise electrical 
measurements. The samples were immersed directly in the 
superfluid helium of the mixing chamber. Electrical connection 
between the millikelvin region and the data acquisition system 
was provided by coaxial cables. Special attention was paid to
control the Joule heating effect. The power dissipation in the samples 
did not exceed $10^{-11}$ W at 100 mK and $10^{-9}$ W at 1 K. 
The estimate of the respective Joule heating, based on the 
Kapitza resistance of the Cu-He boundary $R = 24 $~m$^2$K/W
and $R = 2.4\times  10^{-2}$~m$^2$K/W for 100 mK and 1 K respectively 
\cite{Cu-He} and the energy dissipations quoted above, 
is $\Delta T/T \sim 10^{-1}$.

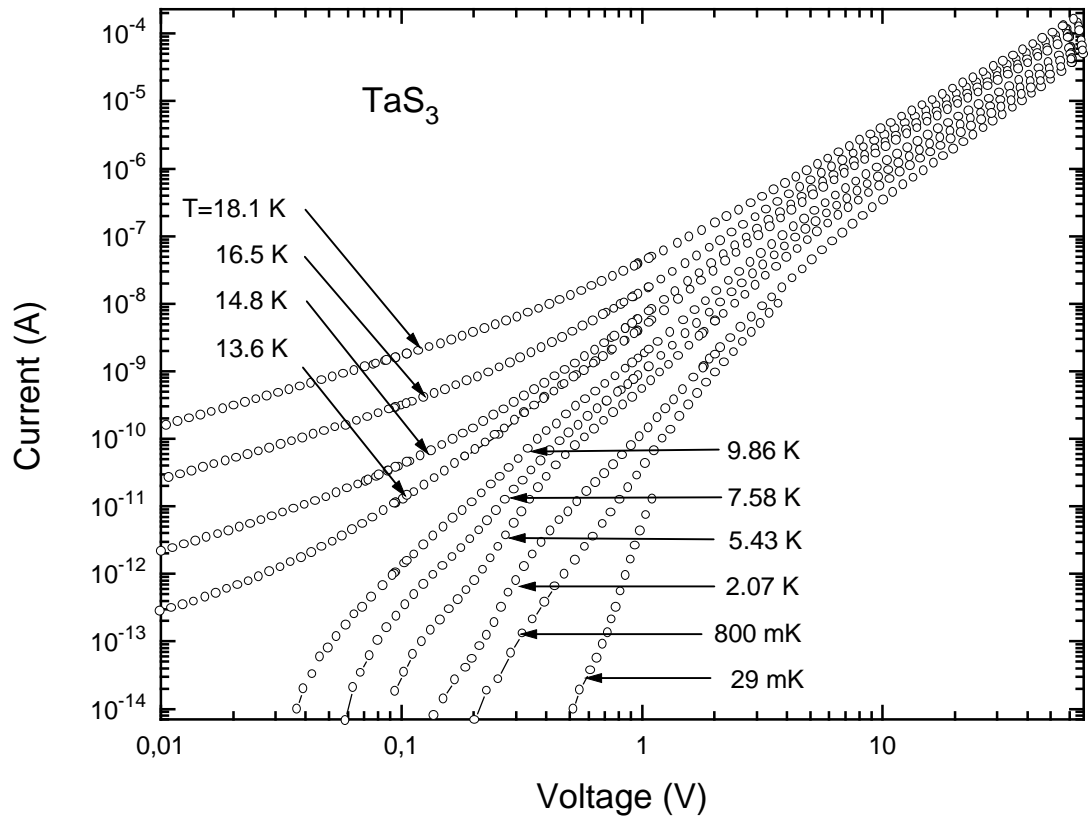
\begin{figure}
\protect
\caption{Current-voltage characteristics of an impure TaS$_3$ single 
crystal in the temperature range 20 K - 30 mK.}
\label{fig:I-V}
\end{figure}

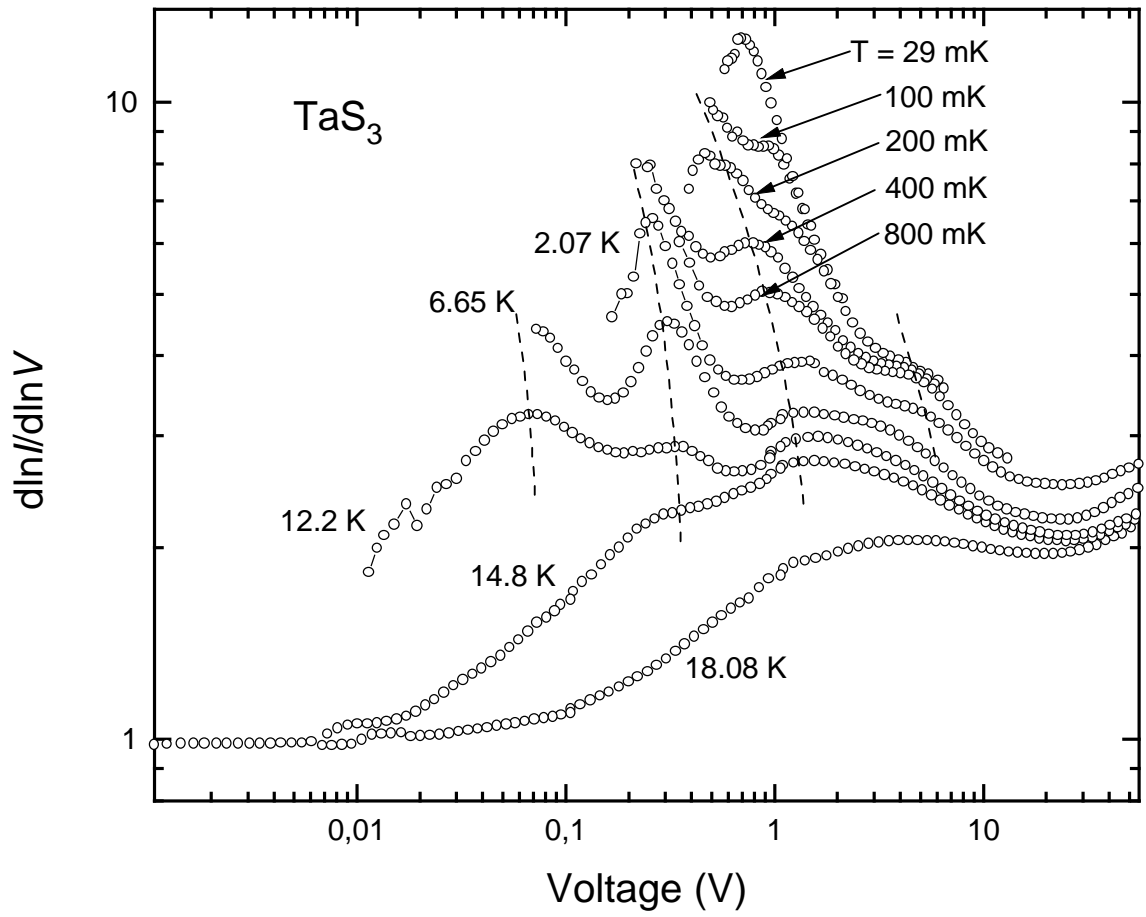
\begin{figure}
\protect
\caption{The logarithmic derivative $\alpha = d\ln I/d \ln V$ 
{\it vs.}
voltage of the same sample as in Fig.~1. Broken lines mark 
the positions of maxima of $\alpha$.}
\label{fig:deriv}
\end{figure}

Figure~\ref{fig:I-V} shows a set of I-V curves taken in the 
temperature range 20 K - 30 mK for an impure TaS$_3$ sample. At 
temperatures above 10 K they have the usual shape reported earlier 
\cite{Tako,INM}
and consist of linear and nonlinear regions. The nonlinear part can 
be roughly approximated by the power law
$I\propto V^\alpha$ with $\alpha$ growing with lowering 
temperature. Figure~\ref{fig:deriv} shows the logarithmic derivative of 
the I-V curves, $\alpha =d\ln I/d\ln V$, of the same sample as in 
Fig.~\ref{fig:I-V}. The linear part corresponds to $\alpha =1$. 

At temperatures below 10 K the current of the linear part is below 
the resolution of the measurements, and only the nonlinear part of I-V 
curve can be observed. A new feature of the nonlinear conduction 
is seen at temperatures around 10 K, where a wave-like 
structure is apparent on the I-V curves (Fig.~\ref{fig:I-V}). The existence of 
this structure is even more evident from Fig.~\ref{fig:deriv}. Up to 
three dips and cusps can be distinguished in $\alpha (V)$ at 
$T < 10$ K. This structure can be attributed to transitions between 
different branches of the I-V curve; the detailed interpretation of the
branches will be given in the discussion. Lowering temperature 
slightly shifts the structure to smaller voltages, whereas the 
respective currents quickly freeze out. The current of the low-voltage 
branches freezes out faster than that of the high-voltage ones.
In other words, the temperature evolution of the I-V curves proceeds 
through sequential freezing-out of the branches.

The fine structure of the I-V curves reported above has been observed 
for all impure samples of TaS$_3$. The current-voltage 
characteristics of pure crystals did not provide such  
unambiguous evidence for the fine structure, though a weak 
two-branch structure may be distinguished at temperatures around 10 K.
 No correlation between the sample sizes and the
shape of the I-V curves was found. Such a dependence on impurity content 
and independence of the geometry proves 
that the fine structure results from 
doping rather than from possible spatial nonuniformity of the
current flow in the highly anisotropic material studied.
 Our analysis of published 
results has shown that similar behavior is also present in other 
CDW conductors, but was overlooked by previous researchers. For 
example, similar wavy I-V curves can be found in published I-V 
curves of m-TaS$_3$ (see Fig.~3 of Ref.~\cite{INM} at $T\leq 31.5$ 
K), though were not mentioned in the respective discussion. We 
conclude therefore that the fine structure of the low-temperature 
I-V curves described above is intrinsic for CDW conductors.


The behavior reported above can be understood in terms of CDW 
creep. In the absence of strong-pinning centers the creeping of the 
CDW occurs in accordance with the general scenario leading to I-V 
curves obeying the equation
\begin{equation}
j=j_0\exp\left[ -\frac{T^*}{T}\left(\frac{E_0}{E}\right) ^\beta\right] ,
\label{eq:exp}
\end{equation}
where $\beta \approx 0.5$ for 3-dimensional pinning \cite{Natterman}. 
Here $j$ is the current density, and $j_0$, $T^*$ and $E_0$ will be
considered below as phenomenological parameters.

Our further discussion will be based on the suggestion that samples of
TaS$_3$ contain impurities with pinning energies high enough to 
provide a strong-pinning contribution \cite{Abe}.
The presence of strong-pinning centers is consistent with the general 
properties of the CDW conductors \cite{Tucker} and agrees with the 
weak-pinning effects observed at $T>100$ K \cite{meso}. In the 
low-temperature region the strong-pinning centers may be
responsible for the temperature-dependent maximum of the 
dielectric constant \cite{Larkin,LB}, for the fast mode of dielectric
relaxation of the CDW \cite{Baklanov}, and for thermodynamic 
anomalies of CDW conductors \cite{Ovchinnikov}.

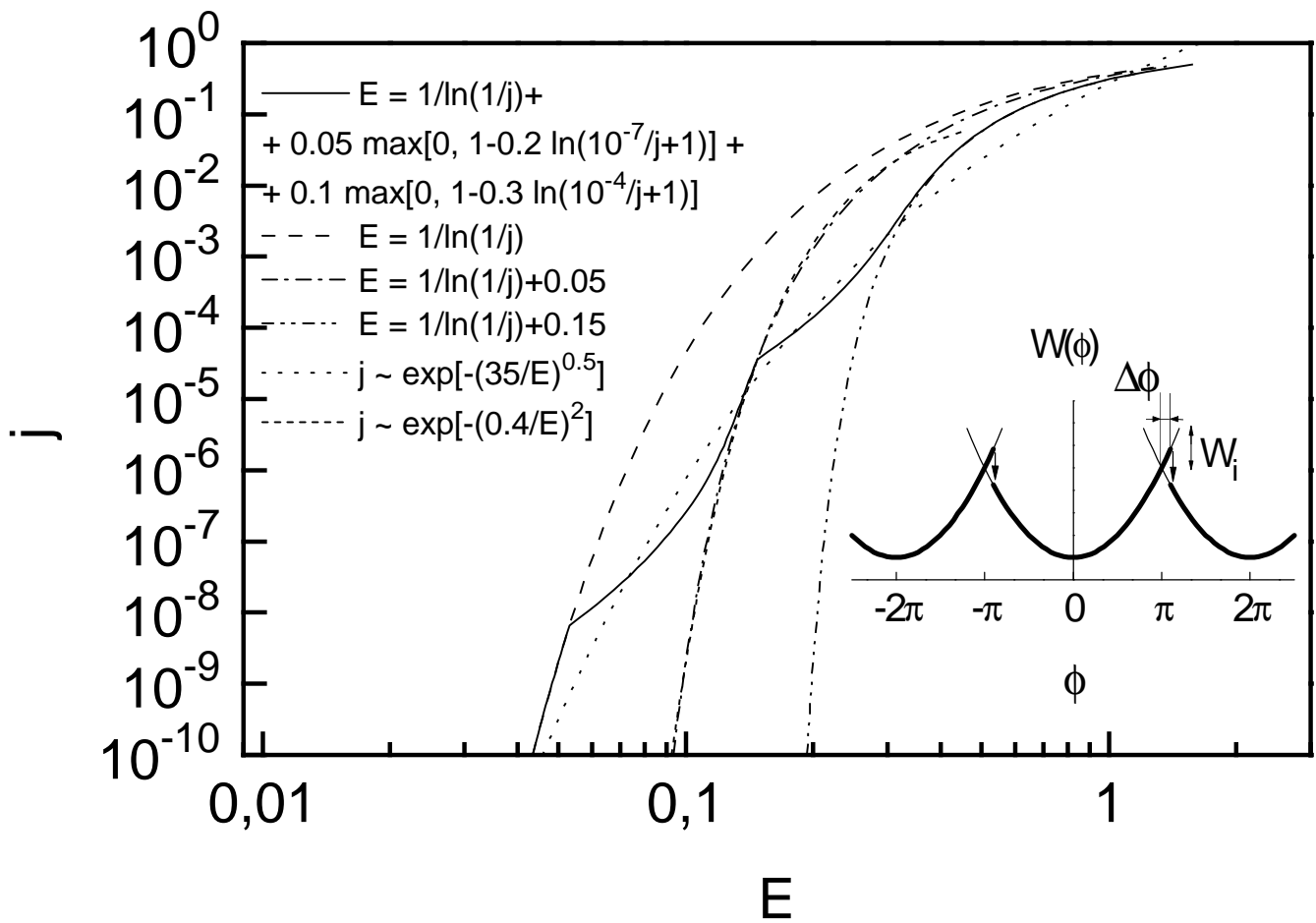
\begin{figure}
\protect
\caption{I-V curve expected in the presence of strong-pinning 
impurities with two different pinning energies (solid line). Dotted 
line corresponds to the I-V curve fit by Eq.~(1) with $\beta = 1/2$. 
Short dash line shows the best fit of $E=1/\ln (j)+0.05$ with
$j\propto \exp [-(E_0/E)^{2}]$.
Inset shows the pinning energy of a strong-pinning impurity 
{\it vs.} 
the CDW phase $\varphi$. The thick line shows the phase 
trajectory of the strong-pinning impurity in the case of stationary 
CDW motion with $d\varphi /dt > 0$ at $T>0$.}
\label{fig:theory}
\end{figure}

Let us initially assume for simplicity that a sample contains only 
strong-pinning impurities of only one sort with pinning 
energy $W_i$ (see inset in Fig.~\ref{fig:theory}) and concentration 
$n_i$. For the CDW at rest ($E=0$) the CDW phases $\varphi$ at the
impurity positions are suggested to be distributed uniformly in the 
interval $-\pi\leq\varphi\leq\pi$ \cite{comment:T=0}. In 
the case of unidirectional creep of the CDW, {\it e.g.} with positive 
$d\varphi /dt$, the phases tend to climb up the upward branch of 
the pinning potential (see inset in Fig.~\ref{fig:theory}). 
 At zero temperature and in the absence of quantum tunneling
the phases reach the upper point of the pinning potential at
$\varphi =\pi + \Delta \varphi_{\max}$ and then
jump to the downward branch. At a finite temperature 
thermal fluctuations allow the barrier to be overcome before reaching the 
upper point. Then the upward part becomes
partially empty, whereas the respective downward branch of
$W(\varphi )$ is partially populated. So the center of the 
distribution is shifted to the right by 
$\Delta\varphi < \Delta\varphi_{\max}$ 
which gives the mean restoring force per one impurity 
$f=f_i\Delta\varphi /\pi$,  where
$f_i\approx (2\pi /\lambda )dW/d\varphi \mid _{\varphi =\pi}$ 
(see inset in Fig.~\ref{fig:theory}). 

Let us introduce the mean time
$t_{2\pi}\equiv e\rho_e \lambda /j$ ($e\rho_e $ is the CDW charge density)
corresponding to a shift of the CDW by its period, $\lambda$, and
the mean time required for activation over the barrier,
$t_{W_i}=t_0\exp\left[ W_i(E)/T\right]$, where $t_0^{-1}$ is the 
respective attempt frequency. Then for $t_{2\pi }\gg t_{W_i}$ the 
barrier $W_i$ is ineffective for pinning, so $f=0$ and the CDW creeps in 
accordance with the general weak-pinning scenario 
\cite{Natterman}. A similar effect of inefficiency of the strong-pinning 
component is known from mesoscopic fluctuations of $E_T$ in small 
samples of TaS$_3$ \cite{meso}.

On the time scale $t_{2\pi }\ll t_{W_i}$
(i.e. at $j\gg j_{W_i}\equiv e\rho_e\lambda /t_{W_i}$) 
\cite{comment:W(E)} the barriers smaller than
$T\ln(t_{2\pi}/t_0)$ can be overcome, and impurities climb up 
to $\Delta \varphi=(1-T\ln(t_{2\pi}/t_0)/W_i)\Delta \varphi_{\max}$. 
So at $j\gg j_{W_i}$ each impurity provides an additional force 
$(1-T\ln(t_{2\pi}/t_0)/W_i)F_i$, where $F_i=f_i\Delta\varphi_{\max}/\pi$. 
The electric field required for the same creep rate is then {\it higher} 
than in the weak-pinning regime by the value
$E_s=(n_i F_i /e\rho_e)\left[1-T\ln(j_{0i}/j+1)/W_i\right]$, where
$j_{0i}= e \rho_e\lambda /t_0$, and $\ln(j_{0i}/j)$ is replaced by 
$\ln(j_{0i}/j+1)$ to include the case $j > j_{0i}$ . Thus the resulting 
I-V curve consists 
of two branches, the weak-pinning branch obeying Eq.~(\ref{eq:exp}), 
and the similar one shifted by $E_s$. The growth of $E$ leads to a 
gradual transition from the weak-pinning branch to the 
strong-pinning one. If impurities provide a variety of pinning
energies $W_i$, then at sufficiently low temperature the I-V curve
consists of several branches as shown schematically in Fig.
\ref{fig:theory} and obeys the equation
\begin{equation}
E=E_0\left( \frac{T}{T^*}\ln\frac{j_0}{j}\right)^{-1/\beta}+%
\sum_{i}\frac{n_i F_{i}}{e\rho_e}\max\left[ 0,\ 1-%
\frac{T}{W_i}\ln\left( \frac{j_{0i}}{j}+1\right)\right] .
\label{eq:strong}
\end{equation}
Equation~(\ref{eq:strong}) gives the extension of Eq.~(\ref{eq:exp}) into the 
strong-pinning area. The strong-pinning effect is therefore
equivalent to the appearance of a {\em series resistance} that
{\em depresses} the
nonlinear conduction \cite{SC} with respect to its background value
provided by
Eq.~(\ref{eq:exp}), and is ineffective at very low creep rate only. 
Note that tunneling through small pinning 
barriers $W_i$ provides an additional channel for nonlinear 
conduction and may contribute to the low-temperature
leveling out of $\log (I)$ {\it vs.} 1/T dependencies well known for
TaS$_3$ below 20 K (see e.g. Refs.~\cite{Tako,INM}). 

The I-V curves of pure TaS$_3$ crystals at temperatures below 1 K can be fitted 
by Eq.~(\ref{eq:exp}) with $\beta =1.5-2$, close to $\beta = 2$ 
reported for thin crystals in the quantum creep regime 
\cite{zzprl}. The general tendency of such a fit is larger $\beta$ 
for purer crystals and lower temperatures. Thus the experimental
value $\beta >1$ is significantly higher than $\beta =0.5$ expected 
for activated creep in the weak-pinning regime \cite{Natterman}. This 
discrepancy may be attributed to quantum phase slip in the bulk 
of a crystal \cite{Ji-Min,Maki,Matsukawa}, 
as well as to the strong-pinning 
effect described above. Note that the presence of strong-pinning 
centers could either increase the exponent $\beta$ 
corresponding to the best fit of I-V curves, or  
decrease it (see Fig. \ref{fig:theory}). 

In accordance with the model described above, the 
low-temperature transport properties of quasi-one dimensional 
conductors depend not only on the impurity concentration, but also 
on the particular impurity type. So a large variety of characteristic
energies and their lack of reproducibility noticed by a number of 
researchers for the low-temperature region (see e.g.
Refs.\cite{review,INM}) may be due to predominance of different 
impurities in the different samples studied.

Summarizing, we observed fine structure in the current-voltage 
characteristics of TaS$_3$ in the CDW creep region. According to 
our model, this structure results from the existence of strong-pinning 
centers with different pinning energies. We 
believe that similar structure can be observed in many other
physical systems where strong pinning centers are present. In 
particular, spin-density waves, vortex lattices in superconductors 
and Wigner crystals are the most probable candidates for observation 
of similar phenomena.

We are grateful to N.~Hegman and G.~Kovacs for technical assistance 
and help during the measurements, K.~Biljakovi\'c, A.~I.~Larkin and 
V.~Ya.~Pokrovskii for useful discussions, and S.~Brazovskii for 
presenting the preprint of his paper. One of us (S.V.Z.-Z.) is very 
grateful to CRTBT-CNRS for kind hospitality during experimental 
research. This work was supported by INTAS (Grant 1010-CT93-0051),
and the Russian Foundation for Basic Research (Grant 95-02-05392).

\begin{references}
\bibitem{nocreep}To be specific, its better to say that  
thermal excitations provide much higher conduction than CDW creep.
\bibitem{comment:T=0} For the sake of simplicity we consider here 
only the case $T\neq 0$ and $T\ll W_i$. In this case metastable 
branches of the pinning potential are not populated.
\end{references}
\end{document}